\documentclass[a4paper,12pt]{article}  
\usepackage[dvips]{epsfig}
\usepackage{amssymb}
\usepackage{cite}
\newcommand{\goes}{\rightarrow} 
 
\newcommand{\GeV}{\; \mathrm{GeV}} 
\newcommand{\TeV}{\; \mathrm{TeV}} 
\newcommand{\beq}{\begin{equation}} 
\newcommand{\eeq}{\end{equation}} 
\newcommand{\bea}{\begin{eqnarray}} 
\newcommand{\eea}{\end{eqnarray}}
\newcommand{\lsp}{\tilde{\chi}}
\newcommand{\mlsp}{m_{\lsp}}

\newcommand{\relic}{\Omega_{\lsp}\,h_0^2} 
 
\newcommand{\etal}{\textit{et al.}}

\begin{document} 
\begin{titlepage} 
 
%%%%%%%%%%% 
\begin{flushright} 
\parbox{6.6cm}{ ACT-12/00, CTP-TAMU-25/00 \\ UA/NPPS-5-00, 
                         hep-ph/0009065 } 
\end{flushright} 
%%%%%%%%%% 
\begin{centering} 
\vspace*{1.5cm} 
 
{\large{\textbf {Neutralino Dark Matter Elastic Scattering
in a Flat and Accelerating Universe}}}\\ 
\vspace{1.4cm} 
 
{\bf A.~B.\ Lahanas} $^{1}$, \, 
{\bf D.~V.~Nanopoulos} $^{2}$  \, and \, {\bf V.~C.~Spanos} $^{1 }$  \\ 
\vspace{.8cm} 
$^{1}$ {\it University of Athens, Physics Department,  
Nuclear and Particle Physics Section,\\  
GR--15771  Athens, Greece}\\ 
 
\vspace{.5cm} 
$^{2}$ {\it Department of Physics,  
         Texas A \& M University, College Station,  
         TX~77843-4242, USA, 
         Astroparticle Physics Group, Houston 
         Advanced Research Center (HARC), Mitchell Campus, 
         Woodlands, TX 77381, USA, and \\ 
         Academy of Athens,  
         Chair of Theoretical Physics,  
         Division of Natural Sciences, 28~Panepistimiou Avenue,  
         Athens 10679, Greece}  \\ 
\end{centering} 
\vspace{3.cm} 
%%%%%%%%%%%%%%%%%%%%%%%%% 
\begin{abstract} 
In SUGRA inspired supersymmetric models with universal boundary conditions for
 the soft masses, the scalar cross section $\sigma_{scalar}$ for the elastic
neutralino--nucleon scattering is in general 
several orders of magnitude
below the sensitivity of current experiments. For large $\tan \beta$
and low $M_{1/2}, m_0$ values, the theoretically predicted $\sigma_{scalar}$
can approach the sensitivity of these experiments ($\approx 10^{-6} \;pb$) being
at the same time in agreement with recent cosmological data, which 
impose severe restrictions on  
the CDM relic density, and with accelerator experiments which
put lower bounds on sparticle and Higgs boson masses. Further improvement of the
sensitivity of DAMA and CDMS experiments will probe the large $\tan \beta$
region of the parameter space in the vicinity of the boundaries 
of the parameter space allowed
by chargino and Higgs searches.
\end{abstract} 
% 

%% \vspace{1.6cm}
%% \noindent
%% \rule[0.cm]{11.6cm}{.009cm} \\     
%% \vspace{.3cm} 
%% $^{\dagger}$ Address after October 2000, 
%% Institut f\"ur 
%% Hochenergiephysik der \"Osterreichischen Akademie
%% der Wissenschaften, A--1050 Vienna, Austria.
\end{titlepage} 
\newpage 
\baselineskip=17.8pt 

%%%%%%%%%%%%%%%%%%%%%%%%%% Letter body %%%%%%%%%%%%%%%%%%%%%%%%%    
As time goes by there is indisputable, accumulative experimental (observational)
evidence that we live in a flat and accelerating universe. Indeed,
the most recent confirmation of such a dramatic statement have been the new
results from the BOOMERANG \cite{boomerang}
 and MAXIMA \cite{maxima} experiments, 
where for the first time, it has been
possible to measure several crucial cosmological parameters simultaneously.
The position of the first acoustic peak of the angular power spectrum
 strongly suggests that
$\Omega_{0}=1$ (i.e. $k=0$, a flat universe), while its shape is consistent with the
predicted inflationary density perturbations. 
It is rather heartwarming to see the way that the data favour an almost
flat Universe $\Omega_{0} \equiv \Omega_{M} + \Omega_{\Lambda}=1 \pm 0.2$
\cite{Turner} where  $\Omega_{M}$ is contribution of the matter density
and $\Omega_{\Lambda}$ is that of the vacuum energy. 
At the same time one determines that $\Omega_{M}= 0.4 \pm 0.1$ which in
conjuction with analysis from high-z SNa data result to
$\Omega_{\Lambda}= \frac{4}{3}\;\Omega_{M} + \frac{1}{3} \pm \frac{1}{6} $
\cite{SNa}. With $\Omega_{M}= 0.4 \pm 0.1$ this implies
that $\Omega_{\Lambda} =0.85\pm 0.2$ \cite{Turner}.
Taking into account the fact that 
the baryonic contribution to the matter density
is small, ${\Omega}_{B}=0.05 \pm 0.005 $,
the values for matter energy density $\Omega_{M}$ 
result to a Cold Dark Matter (CDM) density 
${\Omega}_{\mathrm{CDM}} \simeq 0.35 \pm 0.1$, which combined with 
more recent measurements
of the scaled Hubble parameter
$h_0=0.65 \pm 0.05$,  result to small
CDM relic densities 
$\Omega_{\mathrm{CDM}} \, {h_0}^2 \simeq 0.15 \pm 0.07 $
\cite{Turner}. 
Combined analysis from MAXIMA, BOOMERANG and COBE/DMR  data
entails to even tighter limits \cite{resent}, though we are not using
such stringent limits trying not to extremely constrain the parameter space
of the MSSM.

While the cosmological picture seems to be brighten up at 
least at the level of a phenomenological understanding,
we are facing some squeezing at the other end, i.e. at the fundamental 
particle physics front. The question of
``who ordered this?" for the cosmological constant is so obvious that
needs no further elaboration.
On the other hand the ``observed" value of $\Omega_{M}$ cries out for an
non-baryonic component, rather difficult to be provided by
the nearly massless neutrinos, if we take at face value the present values
of neutrino masses.

In $R$-parity conserving supersymmetric theories the 
lightest supersymmetric particle (LSP or $\lsp$) may be a 
neutralino, which is a good candidate to play the role of  
dark matter (DM) particle.
While it is very encouraging that the neutralino parameters (masses,
couplings) are in the right range to provide easily the right amount
of DM, e.g. the ``correct" contribution to $\Omega_{M}$,
neutralinos still have escaped any direct or indirect detection so far.
The direct way to detect the neutralino is via the neutralino--nucleus
scattering. This scattering contains spin-dependent
as well as spin-independent (scalar) parts.
For the heavy nuclei detectors the spin-independent
part dominates, because there neutron and proton scattering amplitudes
are approximately equal.
The present sensitivity of experiments DAMA \cite{dama} and CDMS \cite{cdms}
measuring the spin-independent
neutralino--proton cross section $\sigma_{scalar}$, is
approximately $10^{-5} - 10^{-6} \;pb$. The CDMS collaboration has reported
negative results from their experiments \cite{cdms}, setting an upper
limit on the spin-independent $\sigma_{scalar}$ that excludes also almost
all the range suggested by the DAMA collaboration \cite{dama}.
Thus in this letter we focus our interest mainly in the CDMS limits.

Since neutralinos seem to play a very important {\em link}
between the very-very small (SUSY theories) and the very-very
large (DM) it is imperative at every stage to have an {\em integrated}
picture of what is allowed and what it isn't.
During the last year we have embarked in such an effort \cite{LNS},
having used cosmological data and electroweak
(EW) precision data (mainly from LEP)
to constrain, in generic SUSY theories, the parameter space.
In this letter we continue our programme \cite{LNS} by injecting
the new cosmological data and new SUSY bounds into our analysis but 
take also into account the most recent CDMS direct-search
exclusion limits.

As we shall see, in general, one has to improve considerably the
sensitivity of the direct-search limits in order to start
excluding sections of the otherwise allowed parameter space.
DM detectors have a current sensitivity
$\approx 10^{-5} - 10^{-6} \;pb$ and would be interesting to find regions
of the MSSM parameter space where bounds imposed by CDMS experiments are
nearly saturated. In most of the parameter space the scalar LSP nucleon
cross section $\sigma_{scalar}$ is small getting values as small as 
$\approx 10^{-9} \;pb$ or even less for LSP masses $m_{\lsp} \approx 200 \GeV$.
This indicates that DAMA and CDMS 
data at the current stage are not that useful to probing MSSM
with universal boundary conditions (UBC) for the soft masses at the
Unification scale. Relaxing the UBC can result to enhancement of the
$\sigma_{scalar}$ as has been studied in Refs \cite{bottino,arno,ellis}.

Our purpose in this paper is to keep on UBC for all soft masses and focus
mainly 
on points that are compatible with neutralino relic densities within the
cosmologically allowed domain  $\relic = 0.15 \pm 0.07$, which yield
cross sections $\sigma_{scalar}$ close to the boundary of the area excluded
by CDMS experiments. For low $\tan \beta < 20 $ this is unlikely to occur
but for higher values cross sections increase with $\tan \beta $ due to the
fact that Higgs bosons become lighter \cite{arno}. It is interesting to point
out that large $\tan \beta $ offers an alternative to neutralino--stau
coannihilation mechanism to decrease neutralino relic densities, in cases
the LSP is a high purity bino. Its small Higgsino component is very important
since it allows it to couple 
with the pseudoscalar Higgs boson  $A$ with a strength that becomes 
sizable for large $\tan \beta$ despite the smallness of the LSP's Higgsino
composition. This in combination with the fact that $A$ mass decreases with
large $\tan \beta$, making pseudoscalar exchanges less suppressive, can lead
to sizable $\lsp \; \lsp \rightarrow b \; {\bar{b}}$  and 
$\lsp \; \lsp \rightarrow \tau \; {\bar{\tau}}$ cross sections and hence low
relic densities. This has been emphasized in Ref. \cite{LNS}. Although
the role of the pseudoscalar $A$ is very significant for this mechanism,
$A$ itself plays no important 
role in  $\sigma_{scalar}$ except that its mass is simultaneously decreased, 
as $\tan \beta$ grows,  
with that of the heaviest of the $CP$-even Higgses which does mediate the
neutralino--nucleon elastic cross section.
Therefore it would be interesting to perform a refined scan of the parameter
space, for large $\tan \beta $, 
to search for points which lead to both small relic densities and values of 
$\sigma_{scalar}$ in the vicinity of the boundary excluded by CDMS data.

In our analysis we assume UBC for the soft masses and focus mainly
to large $\tan \beta $ although results for low $\tan \beta $ will be also
reported. In our evaluation of the scalar cross section 
$\sigma_{scalar}$ we are using the formulae of Ref. \cite{jungman}.
For  the hadronic matrix elements we use the values of the Ref. \cite{ellis}.

Regarding the mass of the pseudoscalar boson which
plays an important role in decreasing the LSP relic density a subtlety arises
related to the theoretical determination of its mass which we should
now discuss.
It is well known that 
radiative corrections to its mass usually calculated through the effective
potential approach are not stable with changing the scale at which they
are calculated. Empirically one calculates its mass at an average
scale $Q_t$ for which 
%%%\approx \sqrt{m_{{\tilde t}_{1}}
the radiative corrections to its mass, due to the
third generation sfemions, are small and hence can be neglected.
In this scheme the pseudoscalar mass squared is given by
$m_{A}^{2} \;=\;-2 \; {{m_{3}^{2}}(Q_t)}/{{\sin 2 \beta} (Q_t) } $, where
$m_{3}^{2} $ is the Higgs mixing parameter. 
 Although this is in principle correct contributions of charginos and
neutralinos are not small at this scale especially 
when $M_{1/2}$ is large. In some case this
may produce an error in the determination of its mass as large as
$25 \%$. Excluding the chargino/neutralino contribution is legitimate
provided the relevant scale is not taken to be $Q_t$, but rather
the average chargino/neutralino mass $Q_{inos}$. At this scale their
contributions can be safely neglected. This scale however may differ
substantially from $Q_t$, when $M_{1/2}$ is large, and third generation
contributions are not small if they are calculated at $Q_{inos}$.
In our approach for the determination of the
pseudoscalar's mass we have duly taken into account all contributions 
including the chargino/neutralino
corrections, as well as the small gauge and Higgs boson contributions,   
and have observed that they  contribute significantly to the
stabilization of the pseudoscalar Higgs boson  mass with changing the scale $Q$.
If these contributions were neglected stabilization would be spoiled
especially in the large $M_{1/2}$ region where the running of the parameter 
${m_{3}^{2}}(Q)$ due to gauginos becomes important and their corrections
should be included to render a $Q$ independent result. 
The situation
would be even more dramatic if in addition to having large $M_{1/2}$  we are
in the large $\tan \beta $ regime where such deviations from
stability are enhanced as being proportional to $\tan \beta $. 
It is also well known that the mass calculated through the effective potential
differs from the pole mass by $\Pi (0) - \Pi (m_A^2)$, where $\Pi (p^2) $
denotes the corrections to the pseudoscalar propagator.
For a more reliable estimation of the pseudoscalar's mass 
we have also included the leading logarithmic
parts of $\Pi (0) - \Pi (m_A^2)$ to the effective potential mass leaving aside
unimportant, at least for our analysis,
small corrections. In this way we have approximated satisfactorily 
the pole mass and avoid the complexities of 
calculating one-loop integrals, which are usually
expressed by Passarino--Veltman functions, as would be demanded if we 
were to determine the pseudoscalar mass through the location of the pole
of the propagator \cite{katkats}.

For consistency, both for the calculation of the pseudoscalar mass and for 
determining the Higgsino and Higgs mixing
parameters $\mu ,{m_{3}^{2}}$ through the one-loop minimization conditions,  
we take into account the contribution of all sectors to the derivatives of
the effective potential not just those of the
third generation.
These are given in Ref. \cite{nath}. Since however they are
given in the Landau gauge, while we are using two-loop RGE's in the 't Hooft's
gauge, we remedy this situation by using the gauge and Higgs boson
contributions to the minimization conditions as these follow directly
from the
tadpole graphs calculated in the 't Hooft's gauge, as can be found in
Ref. \cite{bagger}. More details can be found in Ref. \cite{katkats}. 

Regarding the neutralino relic density
we calculate it in the way prescribed in Ref. \cite{LNS}. 
We solve numerically the Boltzmann transport equation ignoring
at the first stage coannihilation effects \cite{coani},
and in particular 
${\tilde {\tau}} - \lsp$ and in general ${\tilde {l}} - \lsp$ 
coannihilation effects \cite{falk}.
The latter are important for a bino like LSP when its
mass is close to that of the lightest of the ${\tilde \tau}$.
The importance
of these in reducing the relic density has been stressed in Refs
\cite{falk}. One can include the contribution of the coannihilation
processes, whenever they are of relevance, following the empirical rule
%%%%%%%%%
\bea
\Omega_{\lsp} \;=\; R  (\Delta M) \;\Omega_{\lsp}^{0} \,.  \label{red}
\eea
%%%%%%%%%%%%%%%%
The reduction factor $ R (\Delta M)$ depends on
$\Delta M \;=\; (m_{{\tilde {\tau}}_R} - m_{\lsp})/m_{\lsp} $  
and the function $ R (\Delta M)$ smoothly
interpolates between $\approx 0.1$ and 1.0 for values of 
$\Delta M $ in the range $0.00 - 0.25$ (see Ref. \cite{LNS}).
The equation above is a handy device and reproduces 
the results cited in Ref. \cite{falk}.  
$\Omega_{\lsp}^{0}$ appearing on the r.h.s. of the equation above is the
relic density
calculated ignoring coannihilation channels. All effects of coannihilation
processes are effectively included within the function $R$.

As a preview of our results, we have found that the largest 
$\sigma_{scalar}$, approaching values ${10^{-6} - 10^{-7}}\;pb$, are obtained in  
regions of parameter space for which $\tan \beta$ is large.
The dominant contribution to this regime is the Higgs boson  exchange.
For given inputs $m_0, M_{1/2}, A_0$  and the sign of $\mu$, 
Higgs masses decrease as $\tan \beta$ increases. The maximum value of this
angle is determined by the theoretical requirement that we are at the correct
electroweak minimum and Higgs masses squared are positive and bounded
from below by the recent experimental limits \cite{data}.
Hence the contribution of Higgs bosons to 
neutralino--quark elastic cross section becomes
more important in the large $\tan \beta$ regime. Such a decrease in the mass
is not sufficient by itself to increase $\sigma_{scalar}$ to levels
approaching the sensitivity of ongoing experiments. The major role in this
increase plays the coupling of 
the $CP$-even heavy Higgs whose coupling to $d$-quark is proportional to
$\frac{\cos \alpha}{\cos \beta}$.
This is proportional to
$\tan \beta$, when the latter becomes large, since
${\cos \beta} \approx 1/{\tan \beta}$ and the Higgs mixing angle
$\alpha$ becomes small and negative in this region resulting to
$\cos \alpha \approx 1 $.
The situation alters for
the light $CP$-even Higgs boson  whose coupling is $\frac{\sin \alpha}{\cos \beta}$.
In this case  ${\sin \alpha}$ behaves as  $1/{\tan \beta}$ and unlike
the heavy Higgs case its coupling does not grow with increasing
${\tan \beta}$ but stays constant of order unity.
Therefore despite the fact that the heavy $CP$-even Higgs is heavier 
than its light $CP$-even counterpart, its contribution
may be much larger in the large $\tan \beta $  region,  
due to its enhanced coupling to $d$-quark \cite{arno2}. 

From the above argument we conclude that points of the parameter space
that are likely to yield the highest possible elastic cross sections are
those for which ${\tan \beta}$ is large and the heavy $CP$-even state
receives its minimum value allowed by experiments and other theoretical
constraints. The contributions of sfermion exchanges to the amplitude of
the elastic cross sections are less important and will not be discussed.
In the constrained superymmetric scenario  the mass of the $CP$-even heavy state
$m_H$, has the tendency to increase with increasing the effective
supersymmetry scale and therefore low $m_H$ values are obtained for low
$m_0, M_{1/2}$ values. The mass $m_H$ is bounded by the mass of the light
Higgs boson  $m_h$ since
%%%%%%%%%%%%%%%
\beq
m_H^2 \;=\;m_h^2 + \left[ {\frac{1}{4}}  
 \left( (m_A^2 + m_Z^2)^{2} + {\epsilon}^2 \right)
-m_A^2\;m_Z^2\; {(\cos 2 \beta)}^{2} +
{\frac{\epsilon}{2}} {\cos 2 \beta} \;(m_A^2 - m_Z^2) \right] ^{1/2}  \,.   
\eeq
%%%%%%%%%%%%%%%
In this expression $m_A$ refers to the radiatively corrected pseudoscalar
Higgs boson mass and $\epsilon$ are the leading stop corrections to the
$CP$-even Higgs masses. Allowing for additional non-leading contributions,  
or for contributions from other sectors,  
the expression above is modified but for the sake of clarity in presenting our 
arguments we will
use the simplified relation as presented above. In our numerical analysis
we have duly taken into account all stop corrections to the $CP$-even mass
matrix as well as those of sbottoms and staus.

In the
large $\tan \beta$ regime ${\cos 2 \beta} \approx -1$ and the relation above
is simplified to
%%%%%%%%%%%%%%%
\beq
m_H^2 \;=\;m_h^2 + {\frac{1}{4}} (m_A^2 - m_Z^2 - \epsilon)  \,.
\eeq
%%%%%%%%%%%%%%%
Obviously the lowest $m_H$ values are obtained in the region where $m_A$ is
light and $m_h$ is close to its lower experimental bound. The value of
$\epsilon$ can also decrease the mass $m_H$, however this
depends logarithmically on the stop masses and unlike $m_A$ 
does not vary much with changing the SUSY breaking
parameters $m_0, M_{1/2}$ and can be
considered practically stable.
Since Higgs masses increase with increasing the parameters $M_{1/2},m_0$ and
get lowered with increasing $\tan \beta$, the region
of interest the most likely to yield scalar cross sections nearly saturating
the sensitivity of current CDMS experiments is the region of large 
$\tan \beta$ and 
low $M_{1/2}, m_0$ close to the boundaries allowed by Higgs boson and
chargino searches. It becomes evident that increase of the sensitivity of
CDMS experiments will probe this region and may exclude points that would be
otherwise allowed by accelerator experiments.

It is worth pointing out that in the large $\tan \beta$ region 
neutralino relic densities decrease as we have already emphasized
due to both the decrease of the pseudoscalar mass, whose
exchange in $\lsp \lsp \rightarrow b {\bar b} \;,\;\tau {\bar \tau}$
processes is less suppressive, and the
increase of the $\lsp \lsp A$ coupling. 
The smallness of the LSP's 
Higgsino component is compensated by the largeness of $\tan \beta$ yielding
neutralino annihilation cross sections compatible with the recent
astrophysical data when $M_{1/2}$ gets values $<200 \GeV$ \cite{LNS}.
Hence there are regions in which we can obtain both low relic densities
and high $\sigma_{scalar}$ and these regions can be possibly probed
by the next round CDMS experiments. 

\begin{figure}[t] 
\begin{center} 
\epsfig{file=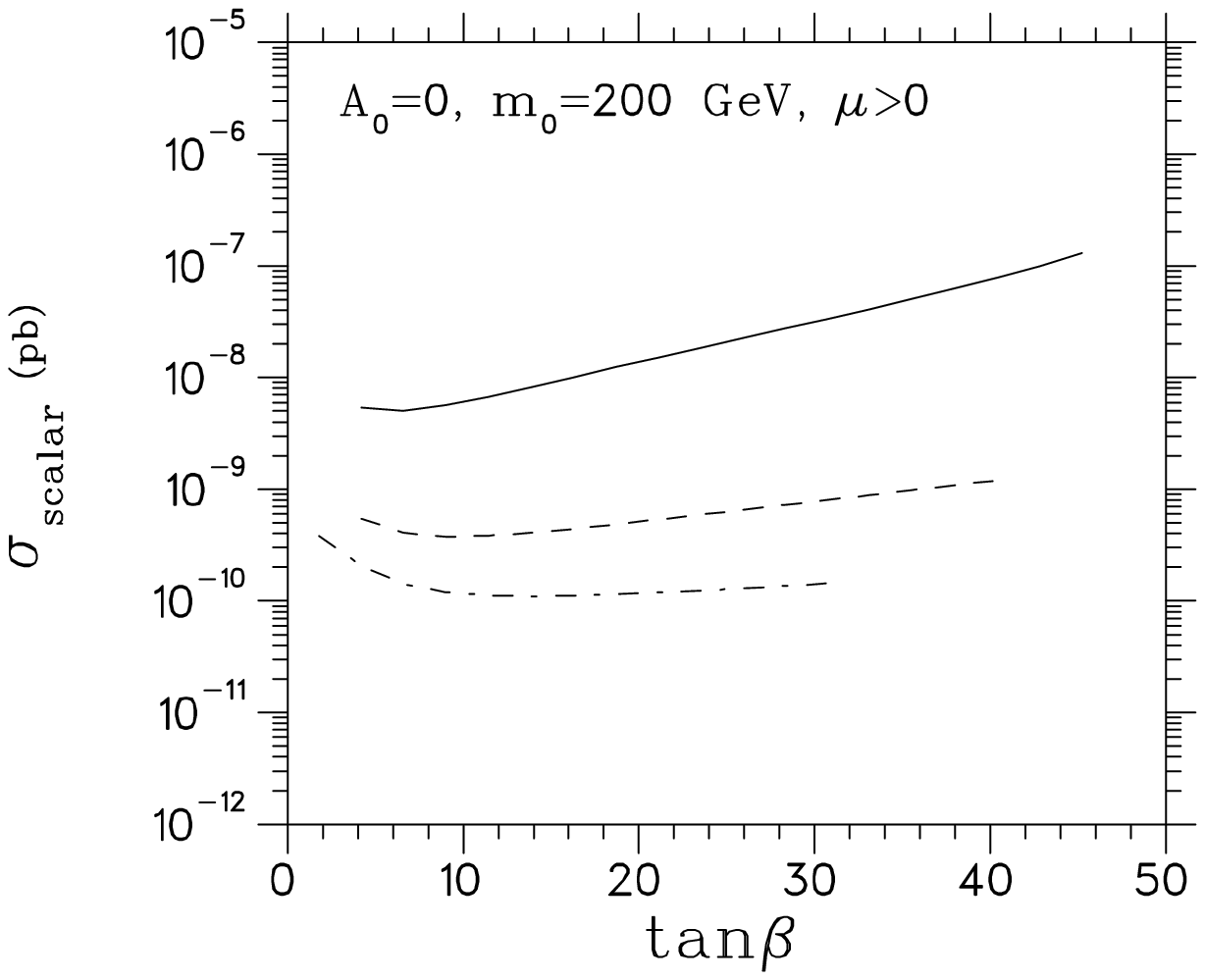,height=9.cm,width=10.cm} 

\vspace{1.cm}
\epsfig{file=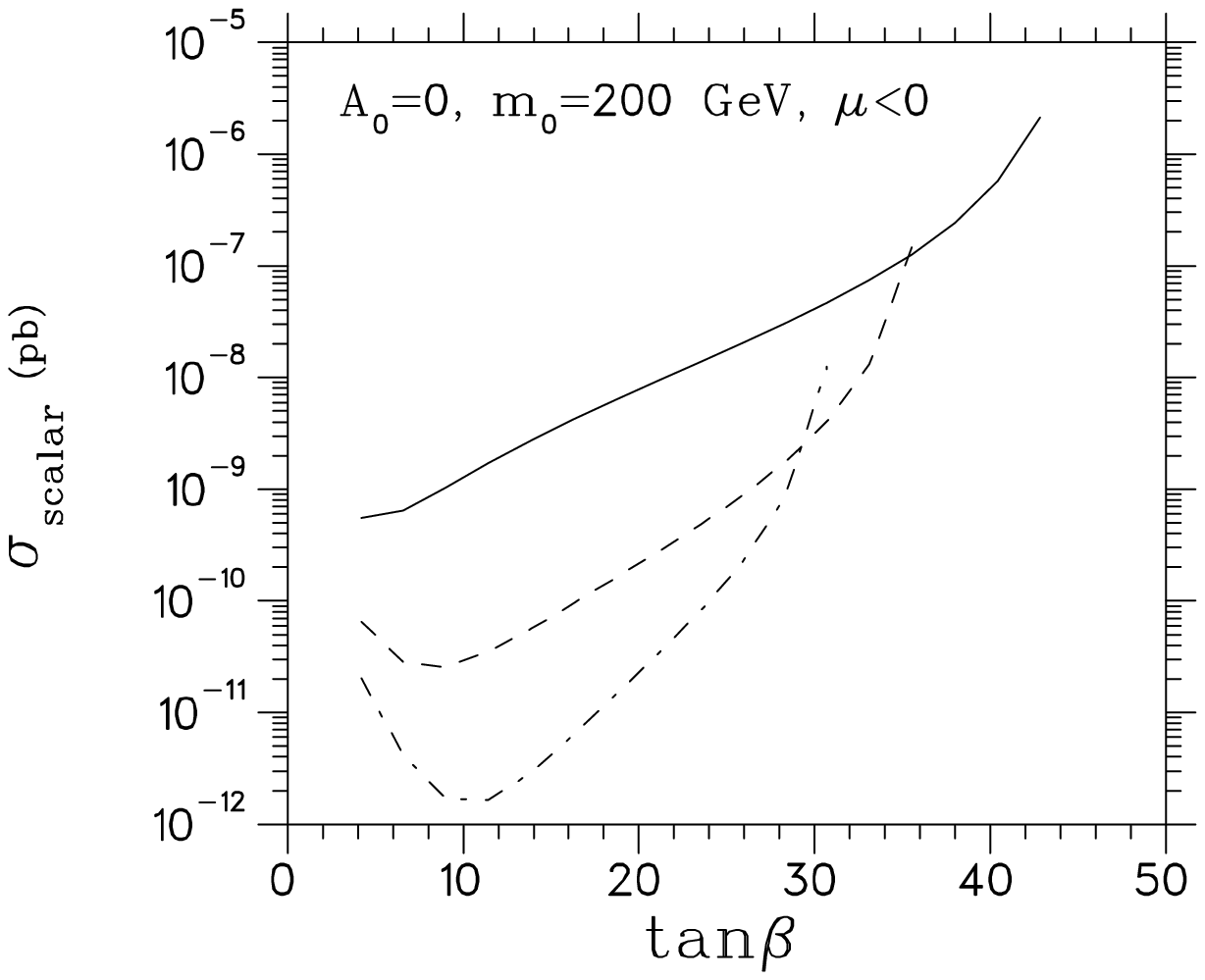,height=9.cm,width=10.cm}
\begin{minipage}[t]{14.cm}  
\caption[]{The $\sigma_{scalar}$, as function of $\tan \beta$,
for both positive
and negative values of the parameter
$\mu$ for given $m_0=200 \GeV, A_0=0 \GeV$. 
The solid, dashed and dot-dashed lines correspond to
$M_{1/2}=200$, $400$ and $600\GeV$ respectively.}

\label{fig1}  
\end{minipage}  
\end{center}  
\end{figure}

For our numerical analysis 
we have scanned regions of parameter space with values
$\tan \beta =1.8 - 50$, $M_{1/2}<1350 \GeV$, $m_0 < 1 \TeV$, $|A_0|< 500 \GeV$.
The dependence on $A_0$ is rather mild and for this reason we cut points 
with values $|A_0| >500 \GeV$. In Figure \ref{fig1} we display the behavior of
$\sigma_{scalar}$ for both positive and negative values of the parameter
$\mu$ for given $m_0=200 \GeV$, $A_0=0 \GeV$ and values of
$M_{1/2}= 200$, $400$ and $600 \GeV$ respectively. One observes that for
$M_{1/2} = 200 \GeV$, which is close to the minimum value allowed by
chargino searches $(\;M_{1/2}\;){{|}_{min}} \;\approx \;170 \GeV$, 
the scalar cross section
increases with increasing $\tan \beta$. Especially in the $\mu < 0$ case this
 increase is more steep and $\sigma_{scalar}$ can reach values slightly above
$10^{-6}\;pb$. In the other cases shown, corresponding to
$M_{1/2} = 400$, $600 \GeV$, the situation is quite different and
$\sigma_{scalar}$ gets smaller by at least an order of magnitude, 
%%stays below $10^{-9}\;pb$
due to the heaviness of all
sectors involved, with the exception of the light Higgs whose contribution 
in this region is however small.
The abrupt stop of all lines towards their right endings
occurs since for higher values of $\tan \beta$ we enter regions which are
theoretically excluded.
%%(absence of correct electroweak vacuum).
Towards the left endings of these lines the light Higgs mass approaches its
lowest experimental bound $\approx 110 \GeV$. For the negative
 $\mu$ case a steep deep is observed for $M_{1/2} =400,600 \GeV$ and 
 $\tan \beta = 10$. This is accidental and due to the cancellation of
sfermion and Higgs contributions to the amplitudes
of the elastic $\lsp$--nucleon cross section \cite{ellis}.

\begin{figure}[t] 
\begin{center} 
\epsfig{file=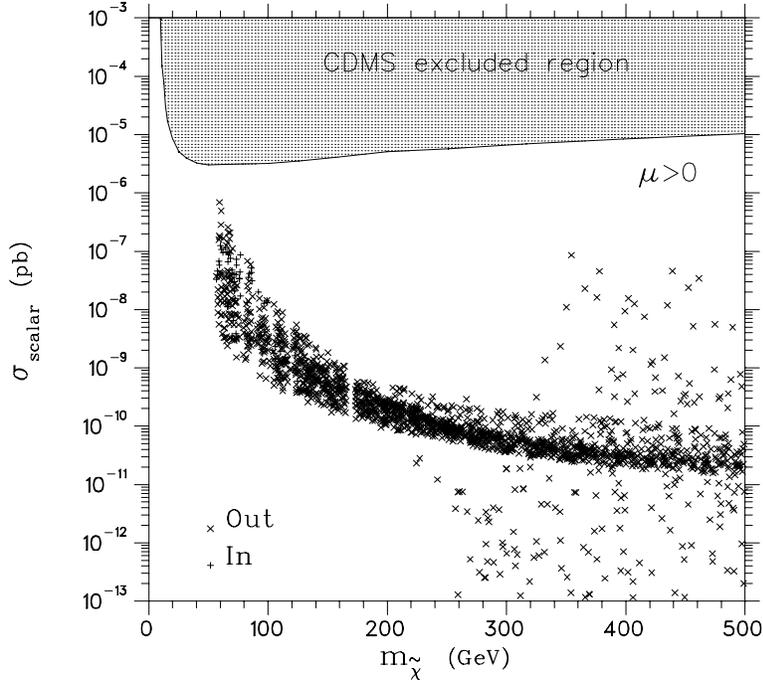,height=9.cm,width=10.cm}
 
\vspace{1.cm}
\epsfig{file=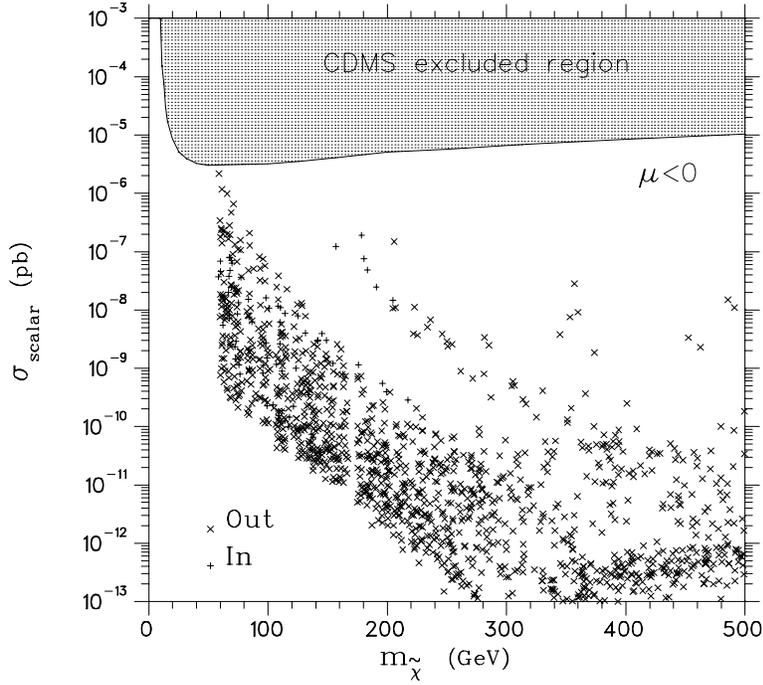,height=9.cm,width=10.cm}
\begin{minipage}[t]{14.cm}  
\caption[]{Scattered plot of the $\sigma_{scalar}$ versus $\mlsp$ from
a sample of 5000 random points in the parameter space.
Both cosmologically acceptable, 
 $\relic=0.15 \pm 0.07$ (marked as ``In''),
and  unacceptable points (marked as ``Out'') are plotted.
Low $\mlsp$ values are excluded by chargino searches.
The shaded region on the top is
excluded by CDMS experiment \cite{cdms}.}
\label{fig2}  
\end{minipage}  
\end{center}  
\end{figure}

In Figure \ref{fig2} we plot the scalar cross section as function of the LSP mass
$m_{\lsp}$. Each point, struck by either a plus or a cross, has been picked
from a sample of 5000 random points in the region of parameter space
mentioned previously. We simultaneously calculate the neutralino relic
density and we denote points which are cosmologically allowed, 
$\relic \;=\;0.15 \pm 0.07$, by a plus and those which are not allowed by a
cross symbol. In this random sample the relic density is calculated without
taking into account the coannihilation processes. We shall return
 to this point later. From this figure it is seen that irrespectively of the
value of the relic density, points that have large cross sections
$\approx 10^{-6} \; pb$ correspond to LSP masses $m_{\lsp} < 80 \GeV$, or
equivalently values of $M_{1/2}$ less than about $200 \GeV$, that is
in the region near the
edge allowed by recent chargino searches. Moreover these points are
characterized
by large $\tan \beta$ ($>30$). Some remarks are in order. The first
concerns the blank stripe occurring at values of $m_{\lsp}$ near $175 \GeV$.
This indicates the presence of a top threshold in
$\lsp \lsp \rightarrow t {\bar t} $ annihilation cross section. This appears
because in order to speed up
calculations we have avoided calculating the relic density at singular  
points, near thresholds
or poles, where non-relativistic expressions usually employed break down, and
more refines techniques must be used to get the correct result. Besides
this we observe the development of other stripes too in the vicinity of
$\approx 100 \GeV$ which correspond to poles of Higgses and other sparticles
involved.

The void region in the negative $\mu$ case for values
$100 \lesssim m_{\lsp} \lesssim 280 \GeV$, 
is due to the fact that for given
value of $\tan \beta, \;m_0$ and $ A_0$, 
an upper bound is set on $M_{1/2}$, or equivalently $m_{\lsp}$, beyond which
we run into situations where EW symmetry is not broken to the
correct vacuum. The larger the value of  
$\tan \beta$ is the lower the value of the upper bound of $M_{1/2}$ is.
Hence points of the sample that yield large cross section for low $M_{1/2}$
values, for values higher than the upper bound set on $M_{1/2}$
are completely absent from the figure since are automatically excluded.
The dispersed points appear for large values of $M_{1/2}$ in the
$\mu >0$ case are related with the behavior of the $\sigma_{scalar}$ as
a function of $\tan\beta$ for large values of $M_{1/2}$.
Actually for such large values of $M_{1/2}$
the light Higgs boson exchange dominates the $\lsp q \goes \lsp q$ process
and therefore the  $\sigma_{scalar}$. However this contribution
is proportional to the $\lsp \lsp h$ coupling, which varies very much
with the  $\tan\beta$ for  large values of $M_{1/2}$, resulting to
the dispersion shown in the figure.

\begin{figure}[t] 
\begin{center} 
\epsfig{file=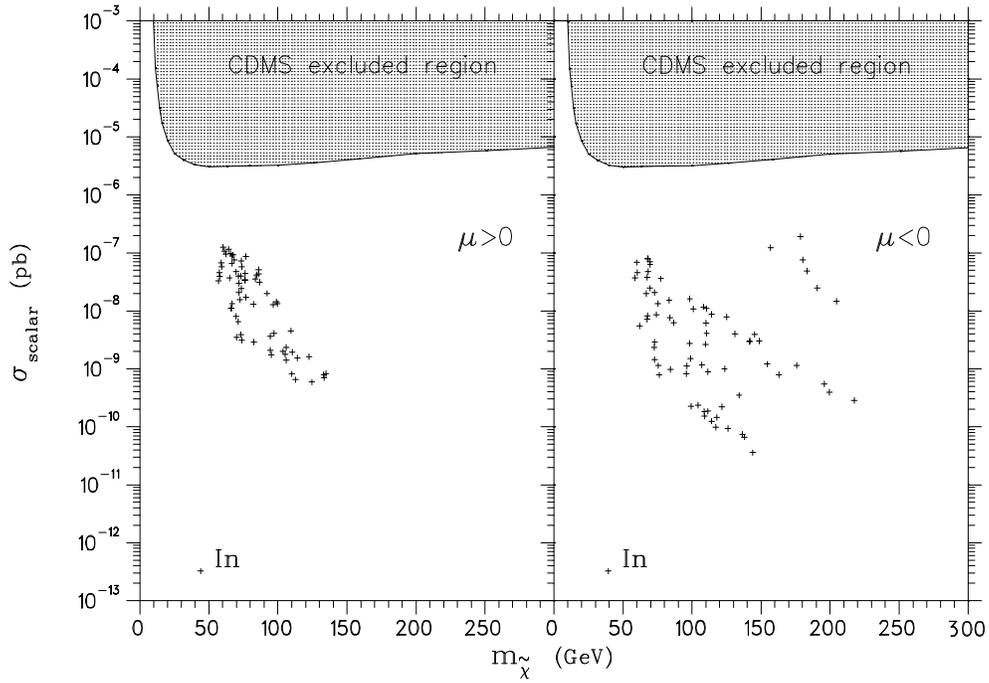,height=9.cm,width=13.cm}
\begin{minipage}[t]{14.cm}  
\caption[]{The same as in Figure \ref{fig2}, where only the cosmologically acceptable
points are plotted. }

\label{fig3}  
\end{minipage}  
\end{center}  
\end{figure}

\begin{figure}[t] 
\begin{center} 
\epsfig{file=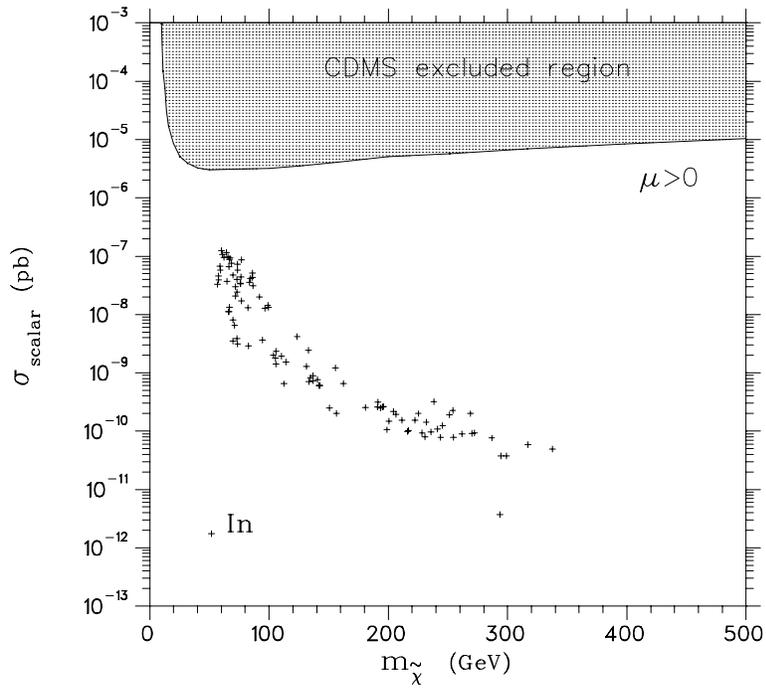,height=9.cm,width=10.cm}
 
\vspace{1.cm}
\epsfig{file=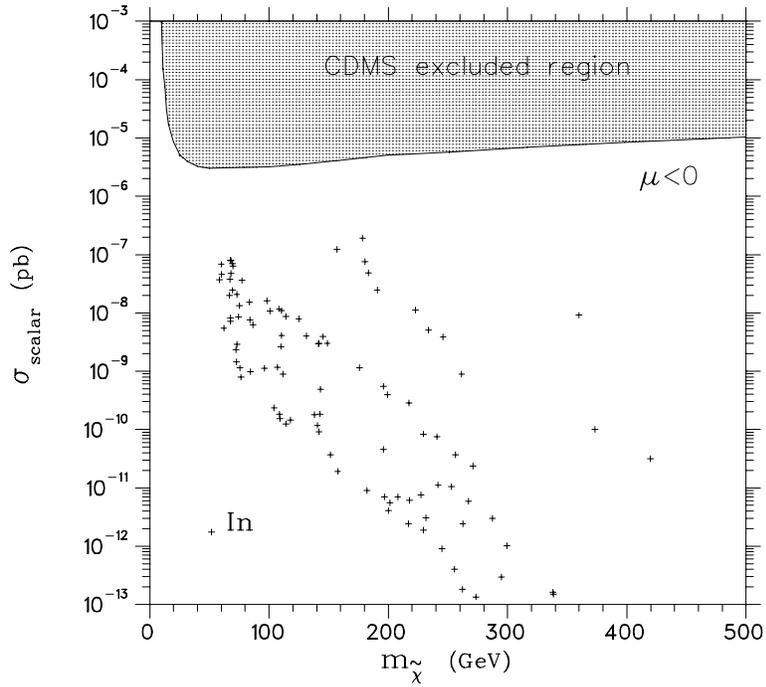,height=9.cm,width=10.cm}
\begin{minipage}[t]{14.cm}  
\caption[]{The same as in Figure \ref{fig3}, where the coannihilation corrections
are taken into account.}

\label{fig4}  
\end{minipage}  
\end{center}  
\end{figure}

Figure \ref{fig3} follows from Figure \ref{fig2} if we discard points that do not fall
within the cosmologically allowed region  $\relic \;=\;0.15 \pm 0.07$. We
observe that only a few points survive for both negative and positive
$\mu$ cases when we enforce this strict cosmological bound on the relic
density. Since we have so far neglected coannihilation processes in our
analysis, and especially ${\tilde \tau} - \lsp$ which are the important
ones when $\lsp$ is mostly a bino, this sample is expected to be enriched when
these processes are taken into account. This is done in Figure \ref{fig4} where
it is shown that the sample of cosmologically allowed points is indeed 
enriched but not in the region which yields the largest
possible cross sections.
The relic density for points that coannihilation processes are of
relevance are calculated as explained in the general discussion earlier in
this paper.

\begin{figure}[t] 
\begin{center} 
\epsfig{file=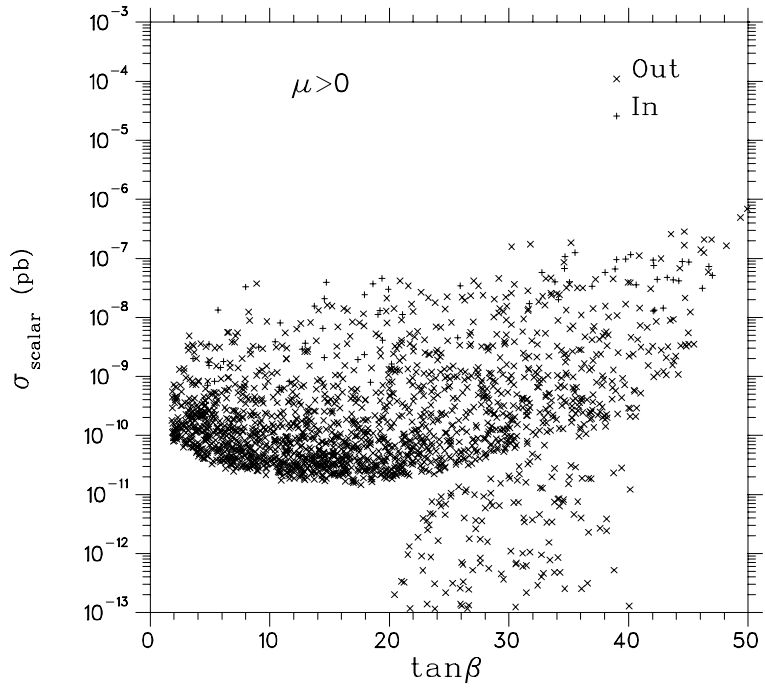,height=9.cm,width=10.cm}
 
\vspace{1.cm}
\epsfig{file=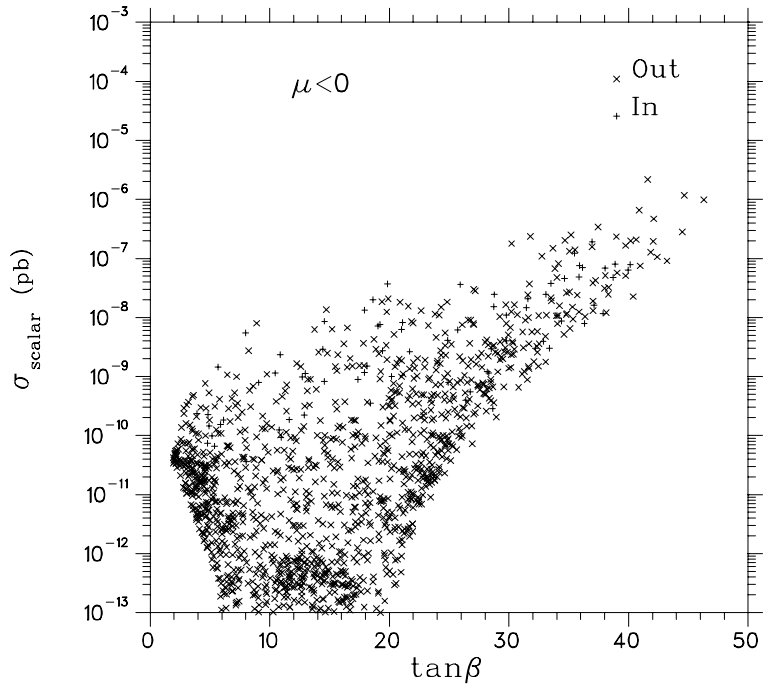,height=9.cm,width=10.cm}
\begin{minipage}[t]{14.cm}  
\caption[]{Scattered plot of the $\sigma_{scalar}$ versus $\tan\beta$ using
the same random sample as in the Figure \ref{fig2}. }
\label{fig5}  
\end{minipage}  
\end{center}  
\end{figure}

In Figures \ref{fig5} we display $\sigma_{scalar}$ as function of $\tan \beta$ for
the random points disused in the previous figures. It is seen that 
points in the large $\tan \beta$ region approach $10^{-6} \;pb$, in the case
$\mu > 0$, and can slightly exceed this value for $\mu < 0$.
These are characterized by large $\tan \beta$ and low $m_0, M_{1/2}$ values
and are in agreement with accelerator experiments on sparticle and Higgs
searches.
Some of these points can survive the strict relic density bounds
indicating that an improvement of the sensitivity of CDMS experiments can
explore the parameter space of MSSM with large $\tan \beta$ and low
$m_0, M_{1/2}$.
Large $\tan \beta$ for $\mu > 0 $ are also compatible with
$b \rightarrow s +\gamma $ data from CLEO \cite{baer},
and hence next run
CDMS experiments may be capable of imposing bounds relevant to regions
of parameter space which accelerator experiments have not probed as yet.

Concluding we have seen that
%%$\sigma_{scalar}$ are no so sensitive to the
%%parameters $A_0$ and $m_0$.
in order to have as large $\sigma_{scalar}$ as possible and close
to CDMS experiment sensitivity, that is $10^{-6} - 10^{-7}$, 
$M_{1/2}$ must be as small as allowed
by chargino searches and $\tan\beta$ as large as possible for the  
Higgs states to be as light as allowed by theoretical constraints and 
experimental searches.
This happens  both  for $\mu > 0$ and  $\mu < 0$.
In addition 
in the  $\mu > 0$ case we can also obtain large 
 $\sigma_{scalar}$ for large  $M_{1/2}\sim 1 \TeV$, again for large
$\tan\beta \gtrsim 30$, 
but these points tend to give unacceptably large
relic densities, $\relic > 1$.
On the other hand for $\mu < 0$ and large $\tan\beta$, we can get large
$\sigma_{scalar}$ for moderate values of  $M_{1/2}\sim 450 \GeV$, 
which are cosmologically acceptable. 
In all cases considered the necessary condition 
in order to have $\sigma_{scalar} \sim 10^{-6} - 10^{-7} $
 is that $\tan\beta \gtrsim 35$, as can be clearly 
seen from Figure \ref{fig5}. It is perhaps important to note that
although the $\mu<0$ case yields, in general, larger cross sections 
than the $\mu>0$ case, the former is disfavoured in view of 
$b \rightarrow s + \gamma $ experimental data \cite{baer}.

\vspace{.4cm}
\noindent 
{\bf Acknowledgments} \\ 
\noindent 
A.B.L. acknowledges support from ERBFMRXCT--960090 TMR programme
and D.V.N. by D.O.E. grant DE-FG03-95-ER-40917.
V.C.S. acknowledges support  
from Academy of Athens under 200/486-2000 grant 
and from European Union under contract HPRN-CT-2000-00149.    
The authors wish to express their thanks to the CDMS collaboration for
 communications.

\newpage
%\vspace{.4cm}
\noindent 
{\bf Note added :} \\ 
\noindent
After submitting this paper for publication we became aware of the paper
hep-ph/0012377 by Bottino, Fornego and Scopel in which values for the scalar
neutralino-nucleon cross section ($\sigma_{scalar}$)
are obtained which can explain the DAMA
data.  The region of the parameter space in this paper in the case of
the mSUGRA model is extended to values of $m_0$ up to $3\TeV$.
In our paper we have taken values for the $m_0$ parameter not exceeding
$1\TeV$. This range is the same with that explored in the paper
hep-ph/0010203 by Bottino, Donato, Fornego and Scopel. The conclusions
reached
in that paper are in agreement with those presented in our work (see
figure 1a of the aforementioned article). Increasing the upper limit of 
$m_0$ to include values  $1\TeV$, we enter regions in which acceptable
neutralino relic densities are obtained with higher values
for $\sigma_{scalar}$, falling within the DAMA experimental sensitivity,
at the cost of having a very massive sfermion spectrum. The region
$m_0>1\TeV$ with $M_{1/2} < 600\GeV$ characterizes the
``Focus Point Supersymmetry" ( see the references
J. Feng, K. Matchev and F. Wilczek, Phys. Lett. B482 (2000) 388;
J. Feng and K. Matchev, Phys. Rev. D63 (2001) 095003). Nevertheless it
should not escape our attention the fact that the ``focus point" region
is now dissalowed by the BNL E821 experiment (H. N. Brown \etal,
{\em{Muon $g-2$ collaboration}}, hep-ex/0102017~) 
and other constraints as emphasized recently
in Ref. hep-ph/0102331 on $g-2$
(J. Ellis, D. V. Nanopoulos and K. Olive, hep-ph/0102331).

\clearpage
%%%%%%%%%%%%%%%%%%%%%%%%%% Figure 1 %%%%%%%%%%%%%%%%%%%%%%%%%%%%%%%% 

\clearpage
%%%%%%%%%%%%%%%%%%%%%%%%%% Figure 2 %%%%%%%%%%%%%%%%%%%%%%%%%%%%%%%% 

\clearpage
%%%%%%%%%%%%%%%%%%%%%%%%%% Figure 3 %%%%%%%%%%%%%%%%%%%%%%%%%%%%%%%% 

\clearpage
%%%%%%%%%%%%%%%%%%%%%%%%%% Figure 4 %%%%%%%%%%%%%%%%%%%%%%%%%%%%%%%% 

\clearpage
%%%%%%%%%%%%%%%%%%%%%%%%%% Figure 5 %%%%%%%%%%%%%%%%%%%%%%%%%%%%%%%% 

\end{document}